\documentclass[acmsmall]{acmart}
\AtBeginDocument{%
  }

\usepackage{algorithmic}
\usepackage{graphicx}
\usepackage{textcomp}
\usepackage{xcolor}
\usepackage{amsmath,amsfonts}
\usepackage{algorithmic}
\usepackage{tcolorbox}
\usepackage{graphicx}
\usepackage{listings}
\usepackage{makecell}
\usepackage{adjustbox}
\usepackage{multirow}
\usepackage{graphicx}
\usepackage{svg}
\usepackage{xspace}
\usepackage{dsfont}
\usepackage{lipsum}
\usepackage{subfigure}
\usepackage{centernot}
\usepackage{diagbox}
\usepackage{tabularx}

\newcolumntype{P}[1]{>{\centering\arraybackslash}p{#1}}
\usepackage[ruled,vlined,linesnumbered]{algorithm2e}

\SetCommentSty{mycommfont}
\usepackage{pifont}
\usepackage{enumitem}

\usepackage[skip=3pt plus 1pt minus 1pt]{caption} 

\newcommand{\tool}{{\texttt{ViBR}}\xspace}

\usepackage{wrapfig}
\usepackage{xcolor}
\usepackage{xcolor,colortbl}
\definecolor{lightgray}{gray}{0.93}
\definecolor{slightgray}{gray}{0.98}
\definecolor{darkgray}{gray}{0.77}

\definecolor{amber}{rgb}{1.0, 0.49, 0.0}

\usepackage{framed}
\definecolor{formalshade}{rgb}{0.95, 0.95, 1}
\definecolor{mygray}{gray}{0.4}
\definecolor{lightgray}{gray}{0.93}

\usepackage[normalem]{ulem} 

\setcopyright{rightsretained}
\acmDOI{10.1145/3808152}
\acmYear{2026}
\acmJournal{PACMSE}
\acmVolume{3}
\acmNumber{FSE}
\acmArticle{FSE145}
\acmMonth{7}
\acmSubmissionID{fse26mainb-p692-p}
\received{2026-02-24}
\received[accepted]{2026-03-24}

\begin{document}

\title{ViBR: Automated Bug Replay from Video-based Reports using Vision-Language Models}


\author{Sidong Feng}
\orcid{0000-0001-7740-0377}
\affiliation{%
  \institution{The Chinese University of Hong Kong (Shenzhen)}
  \city{Shenzhen}
  \country{China}
}
\email{sidongfeng@cuhk.edu.cn}

\author{Dingbang Wang}
\orcid{0009-0002-9675-6824}
\affiliation{%
  \institution{University of Connecticut}
  \city{Storrs}
  \country{USA}
}
\email{dingbang.wang@uconn.edu}

\author{Nikola Tomic}
\orcid{0009-0002-4840-7240}
\affiliation{%
  \institution{TU Munich}
  \city{Heilbronn}
  \country{Germany}
}
\email{nikola.tomic@tum.de}

\author{Tingting Yu}
\orcid{0000-0002-9461-4251}
\affiliation{%
  \institution{University of Connecticut}
  \city{Storrs}
  \country{USA}
}
\email{tingting.yu@uconn.edu}

\author{Aldeida Aleti}
\orcid{0000-0002-1716-690X}
\affiliation{%
  \institution{Monash University}
  \city{Melbourne}
  \country{Australia}
}
\email{aldeida.aleti@monash.edu}

\author{Chunyang Chen}
\orcid{0000-0003-2011-9618}
\affiliation{%
  \institution{TU Munich}
  \city{Heilbronn}
  \country{Germany}
}
\affiliation{%
  \institution{Monash University}
  \city{Melbourne}
  \country{Australia}
}
\email{chun-yang.chen@tum.de}


\renewcommand{\shortauthors}{Feng et al.}

\begin{abstract}
  Bug reports play a critical role in software maintenance by helping users convey encountered issues to developers. Recently, GUI screen capture videos have gained popularity as a bug reporting artifact due to their ease of use and ability to retain rich contextual information. However, automatically reproducing bugs from such recordings remains a significant challenge. Existing methods often rely on fragile image-processing heuristics, explicit touch indicators, or pre-constructed UI transition graphs, which require non-trivial instrumentation and app-specific setup. This paper presents \tool, a lightweight and fully automated approach that reproduces bugs directly from GUI recordings. Specifically, \tool combines CLIP-based embedding similarity for action boundary segmentation with Vision-Language Models (VLMs) for region-aware GUI state comparison and guided bug replay. Experimental results show that \tool successfully reproduces 72\% of bug recordings, significantly outperforming state-of-the-art baselines and ablation variants.
\end{abstract}



\maketitle

\section{Introduction}
\label{sec:introduction}
Handling bug reports is a fundamental task in software maintenance, with bug reproduction often being the first and most critical step toward effective localization and resolution. 
However, many bugs are encountered by non-technical users who may lack the expertise or motivation to craft clear, structured reports~\cite{aranda2009secret, bettenburg2008extracting}. 
Poorly written reports are frequently misunderstood by developers~\cite{feng2024enabling,feng2025breaking,feng2025agent}, leading to repetitive communication and delays in testing and fixing the issue.


To lower the barrier to bug reporting, video-based bug recordings~\cite{feng2022gifdroid}—screen capture videos of issues—have gained traction as a more accessible and expressive alternative to traditional reports. 
This trend is also reflected in platforms like GitHub, which now support video uploads in issue reports to foster clearer communication and faster collaboration between users and developers~\cite{web:GitHubVideoUpload}.
First, it is easy to record the screen as there are many tools available~\cite{web:BugClipper, web:TestFairy}, some of which are even embedded in the operating system by default like iOS~\cite{web:iosrecord} and Android~\cite{web:androidrecord}.
Second, GUI recording can retain rich runtime context such as environmental configurations, dynamic content, and user input parameters, hence it bridges the understanding gap between users and developers.


\begin{wrapfigure}{r}{0.45\linewidth}
\centering 
	\includegraphics[width=\linewidth]{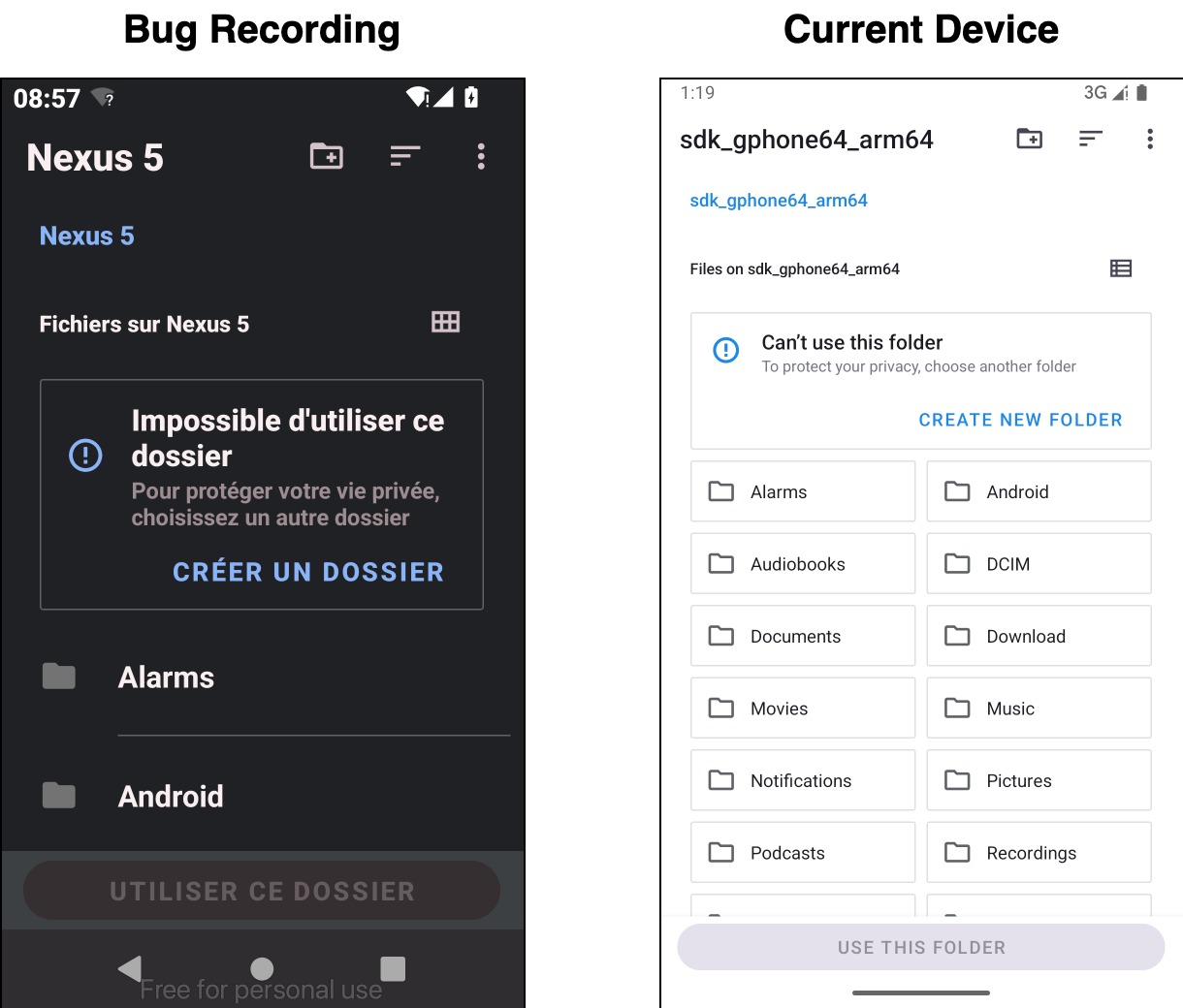} 
	\caption{GUI comparison between recording and device.} 
	\label{fig:difference}
\end{wrapfigure}

Despite their growing prevalence, reproducing bugs from GUI recordings remains a manual and error-prone process.  
Developers must watch the raw footage to infer user actions and the involved GUI elements, which is often ambiguous and time-consuming. This process becomes even harder due to cross-device inconsistencies, where the same GUI functionality may appear with different layouts, resolutions, themes, or widget placements across devices, making it difficult to reliably map recorded steps to the current device state. For example, in Fig.~\ref{fig:difference}, the same folder-selection interface can vary substantially across devices and configurations, such as 1920$\times$1080 vs. 1440$\times$2560 resolutions, list vs. grid layouts, dark vs. light themes, and French vs. English language settings.
Even small variations can hinder faithful bug reproduction on test devices.

While many existing techniques target bug reproduction from textual reports using pattern analysis and even state-of-the-art large language models (LLMs)~\cite{zhao2019recdroid,fazzini2018automatically,wang2024feedback,feng2024prompting}, these methods are not applicable to video-based reports, which lack structured step-by-step descriptions.
Some recent efforts have explored video-based replay. 
For example, GIFdroid~\cite{feng2022gifdroid} aligns keyframes from GUI recordings to states in a pre-constructed UI Transition Graph (UTG) using pixel-level matching. 
However, this approach is brittle in real-world scenarios, where layout changes or dynamic GUI elements can easily break the alignment. 
V2S~\cite{bernal2020translating} leverages deep learning to detect touch indicators within the recording and replays those interactions accordingly. 
Yet, it requires users to enable touch indicators in advance when recording bugs. In our analysis of real-world GUI recordings, only 21.5\% included visible touch indicators, which imposes a heavy setup burden and limits its practicality for widespread adoption.
Consequently, developing a lightweight, automated, and robust approach for replaying bugs from GUI recordings remains a critical need.

In this paper, we present \tool, a novel framework for automatically performing \textbf{Vi}deo-based \textbf{B}ug \textbf{R}eplay without requiring app instrumentation, touch indicators, or pre-built UI graphs.
Inspired by recent advances in Vision-Language Models (VLMs), like state-of-the-art GPT-4o, \tool formulates bug reproduction as a multi-modal reasoning problem: given a recording and the current GUI state, the approach infers the next actionable step and executes it accordingly.
Specifically, \tool (1) segments the recording into distinct user interaction scenes by comparing CLIP embeddings of consecutive frames; 
(2) employs region-aware VLM-based reasoning to compare functionally equivalent GUI states between the recording and the target device;
and (3) adaptively infers and replays the corresponding user actions on the device.
Unlike prior approaches, \tool analyzes GUI recordings in a global-to-local manner: it first identifies interaction boundaries at the video level, and then performs fine-grained region-level reasoning to locate and operate on the correct GUI elements on the target device.


We conduct comprehensive evaluations of \tool across the three phases of our approach.
First, we evaluate our approach in segmenting the action scenes from 75 GUI recordings that are widely-studied in the prior studies, achieving up to 87\%, 85\%, and 86\% in precision, recall, and F1-score, respectively.
Second, we assess our approach in identifying functional consistency between the recorded and current GUI states.
Compared with four state-of-the-art baselines and two ablation study, our approach achieves significantly better performance, i.e., 86\% in precision, 88\% in recall, and 87\% in F1-score.
In addition, we evaluate the \tool in guiding bug replay and the results show that our approach can successfully reproduce 72.0\% from real-world bug recordings with minimal runtime overhead (less than a few cents), significantly improving over existing methods.
The contributions of this paper are as follows:
\begin{itemize}
    \item To the best of our knowledge, this is the first work that leverages Vision-Language Models for cross-device GUI-state consistency reasoning over GUI recordings, opening new possibilities for GUI-related engineering tasks.
    \item We introduce \tool, a lightweight, fully automated framework that reproduces bugs directly from GUI recordings without requiring any heavy setups.
    \item We perform a comprehensive evaluation on GUI recordings, demonstrating the effectiveness of \tool compared to state-of-the-art baselines and ablation studies.
\end{itemize}


\begin{figure*}
	\centering 
	\includegraphics[width=0.85\textwidth]{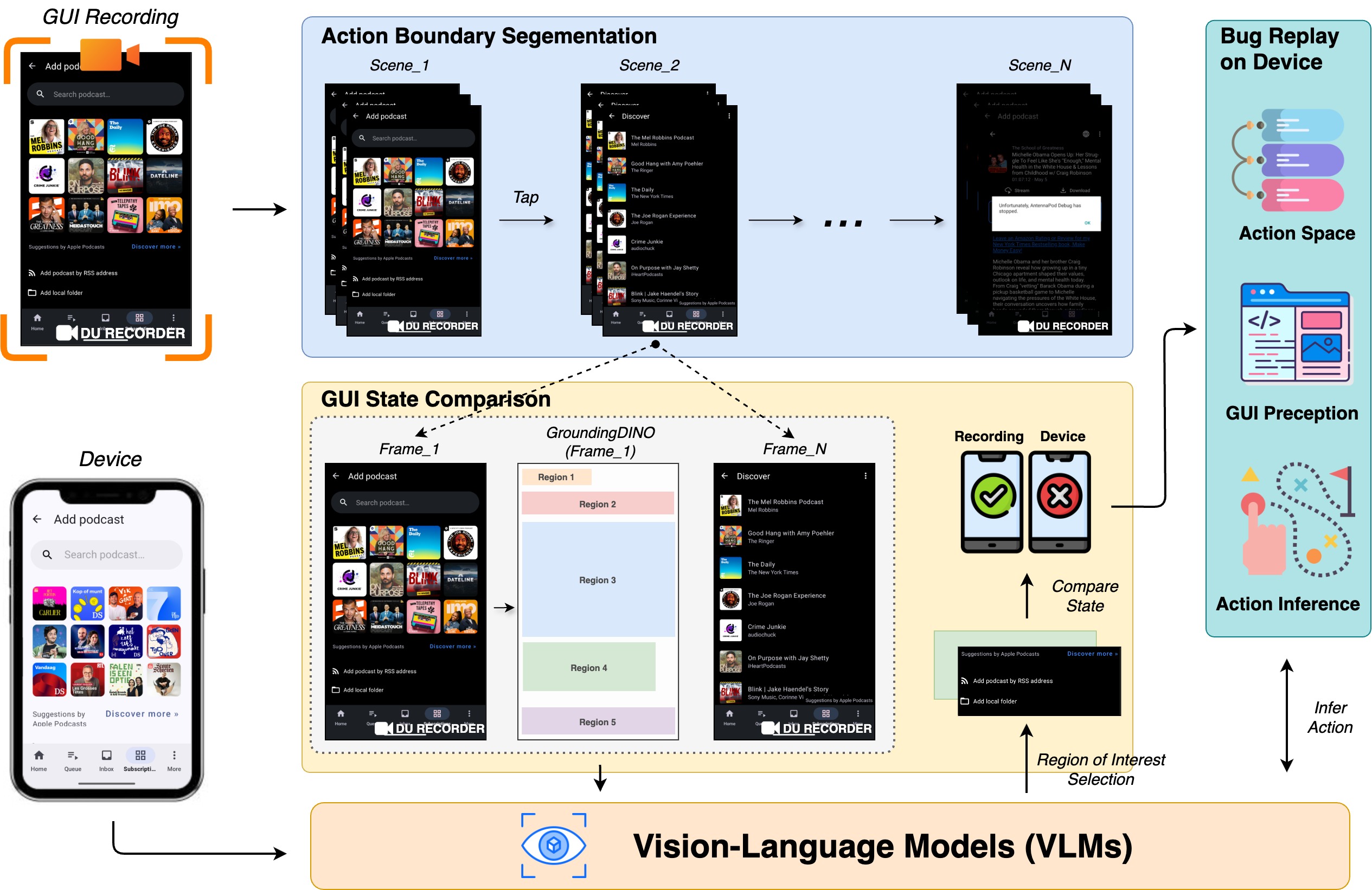} 
	\caption{The overview of \tool.} 
	\label{fig:overview}
\end{figure*}

\section{\tool Approach}


We present an automated approach for reproducing bugs from GUI recordings by segmenting the input video into user interaction scenes and conditionally replaying each action based on the current GUI state of the device. 
An overview of our approach is shown in Fig.~\ref{fig:overview} and comprises three key phases:
(i) the \textit{Action Boundary Segmentation} phase, where the GUI recording is divided into distinct scenes, each representing a single user action;
(ii) the \textit{GUI State Comparison} phase, which determines whether the GUI state associated with each scene matches the current screen on the device; and
(iii) the \textit{Bug Replay on Device} phase, which infers the intended user action and adaptively executes it on the device to reproduce the bug.



\subsection{Action Boundary Segmentation}
\label{sec:phase1}
While recent work~\cite{zhao2023learning,weng2024longvlm} has explored Vision–Language Models (VLMs) for natural video segmentation, these models are not tailored to domain-specific video types such as GUI recordings, where transitions are driven not by natural scene changes but by discrete user interactions.
To bridge this gap, we adopt a signal-processing perspective, treating GUI recordings as ordered sequences of frames that encode state transitions initiated by user actions. 
Unlike natural-scene segmentation, which primarily focuses on detecting broad visual discontinuities, segmenting GUI recordings requires capturing fine-grained, user-driven state changes.
Such transitions are often localized, triggered by actions like clicks, scrolling, or text input, yet they result in significant shifts in the interface layout. 
According to this insight, we segment GUI recordings into discrete user action scenes by analyzing the visual similarity between consecutive frames.

\begin{figure}
	\centering 
	\includegraphics[width=0.99\linewidth]{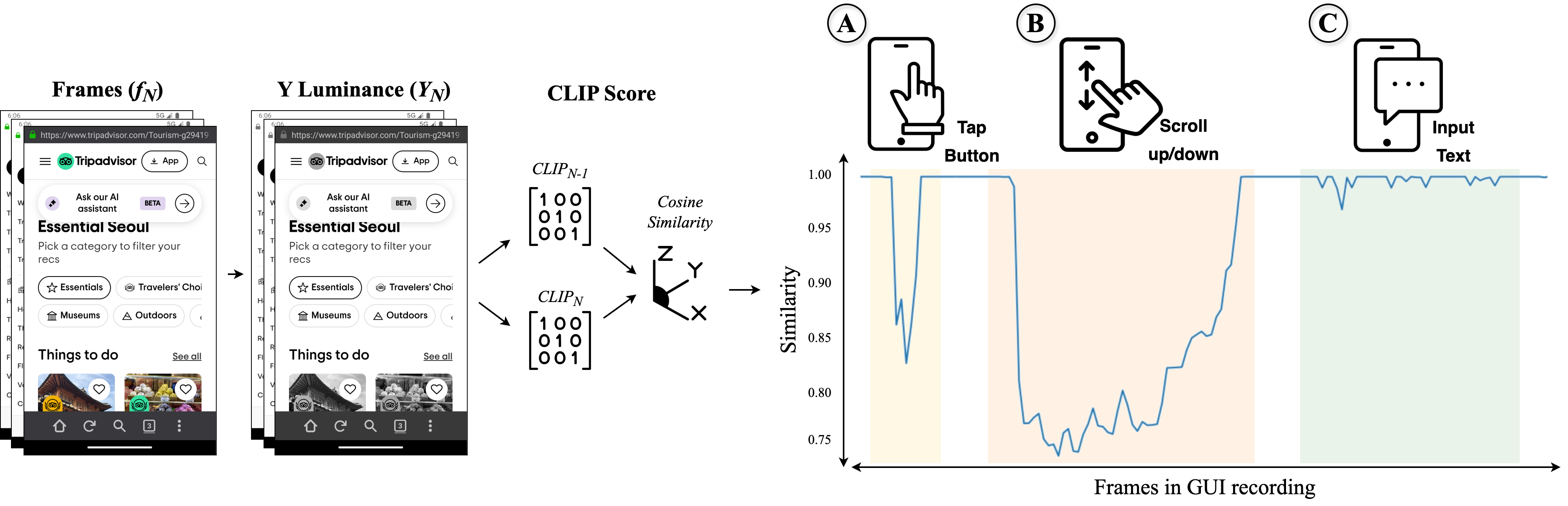} 
	\caption{An illustration of consecutive frame similarity.} 
	\label{fig:timeframe}
\end{figure}

\subsubsection{Consecutive Frame Comparison}
Consider a GUI recording $\big\{ f_{0}, f_{1}, .., f_{N-1}, f_{N} \big\}$, where $f_{N}$ is the current frame and $f_{N-1}$ is the previous frame, as shown in Fig.~\ref{fig:timeframe}.
Since raw video streams often contain visual noise (such as compression artifacts, color fluctuations, or subtle pixel jitter from screen refresh), we first reduce irrelevant variations by adopting a luminance-based representation.
Inspired by a prior work on video segmentation~\cite{feng2022gifdroid}, we employ Y-Difference (Y-Diff), a metric derived from the Y channel of the YUV color space, which is widely used in video compression due to its alignment with human visual perception—particularly the Y (luminance) component that dominates the recognition of structural and motion-related changes~\cite{chen2009compression,sudhir2011efficient,feng2022gifdroid}.
In practice, we convert each frame from RGB to YUV and extract the luminance mask $Y_{N}$ for frame $f_{N}$.

Next, to capture both semantic and structural information of the Y-Diff frame, we leverage the lightweight semantic embedding model CLIP (Contrastive Language–Image Pre-training)~\cite{radford2021learning}.
We select CLIP because, unlike handcrafted similarity metrics that are highly sensitive to low-level perturbations (e.g., small shifts or noise), CLIP is trained on large-scale image–text pairs to learn a joint embedding space.
This enables the embeddings to remain robust to such variations while encoding high-level perceptual and semantic information, yet still preserving meaningful pixel-level differences. 
Specifically, the luminance representations $Y_{N}$ are first normalized and then passed through the CLIP image encoder, which produces high-dimensional embeddings $CLIP_{N}$ that preserve both the structural layout and the semantic content of the frame.

Finally, we compute the similarity between consecutive frames by applying the widely-used cosine similarity metric between their CLIP embeddings.
The resulting similarity score is normalized between 0 and 1, where a higher value indicates strong visual and semantic similarity, while a lower value reflects substantial changes, typically corresponding to a GUI transition triggered by user interaction.

\subsubsection{Frame Grouping}
\label{sec:action_class}
After computing pairwise CLIP scores, we group frames into atomic activity, e.g., user action, following prior work~\cite{feng2022gifdroid}.
This procedure is essential because discrete user activities typically span multiple frames, requiring them to be grouped into coherent scenes that reflect each discrete action.
To achieve this, we analyze the similarity series as shown in Fig.~\ref{fig:timeframe} to identify coherent segments.
Consequently, we identify patterns for three primary interaction types: \textit{Tap}, \textit{Scroll}, and \textit{Input}. 
Note that we focus on the most commonly-used actions for brevity in this paper, other actions could be extended by comparing the consecutive frame similarity.

\textit{(a) Tap}: 
Characterized by a sharp drop in similarity as shown in Fig.~\ref{fig:timeframe}-A, indicating a fast transition to a new screen, often triggered by clicking a button.

\textit{(b) Scroll}: 
Typically begin with an initial drop, followed by gradual increases in similarity as shown in Fig.~\ref{fig:timeframe}-B, reflecting the continuous nature of vertical scrolling where content shifts incrementally.

\textit{(c) Input}:
Involve oscillations in the similarity score as shown in Fig.~\ref{fig:timeframe}-C, as the GUI reflects incremental changes while characters are typed. 
However, distinguishing input actions purely based on similarity is unreliable due to potential overlaps with tap-like transitions (e.g., focusing on a text field). 
Therefore, we incorporate Optical Character Recognition (OCR) to explicitly detect the appearance of virtual keyboards. Specifically, we adopt the state-of-the-art PPOCR~\cite{du2021pp} as our OCR model. 
For each frame, we extract OCR-detected characters and split them into two sequences: $ocr_{text}$ for alphabetic content and $ocr_{num}$ for numeric content.
Given OCR's imperfections in low-resolution or occluded text regions, we rely on the presence of keyboard-specific substrings to identify keyboard frames.
For example, the detection of substrings such as \textit{``qwert''} in $ocr_{text}$ is indicative of an English keyboard, while patterns like \textit{``123''} in $ocr_{num}$ suggest a numeric keypad. 
We formally define a keyboard frame as one that satisfies:
\begin{equation}
	frame = 
	\begin{cases}
		\exists \text{\{qwert, asdfg, zxcvb\}} \in lowercase(ocr_{text}) \\
		\exists \text{\{123, 456, 789\}} \in ocr_{num} \\
	\end{cases}
	\label{eq:keyboard}
\end{equation}
We deliberately avoid template-based keyboard matching, as the appearance of virtual keyboards can vary significantly across devices, due to custom themes, screen sizes, and input methods.


\subsection{GUI State Comparison}
\label{sec:phase2}
GUIs are inherently dynamic. Variations such as pop-up dialogs, layout shifts, and overlay configurations can introduce inconsistencies between the recorded environment and the current runtime state. 
These discrepancies often hinder accurate bug reproduction, as action scenes may not align with the live application interface, even on the same device.
Therefore, we aim to verify whether the current GUI state matches the recorded scene before executing the corresponding user action.

A seemingly straightforward solution would be to apply traditional image similarity metrics to compare the recorded frame with the current screen. 
However, such methods fail to capture semantic equivalence. 
Visually distinct screens may support identical interactions (e.g., tapping ``Android'' folder button in Fig.~\ref{fig:difference}), while visually similar screens may differ in functionality. 
This semantic gap renders pixel-level techniques insufficient for GUI reasoning.

To address this, we propose a region-guided, attention-based comparison framework that leverages Vision-Language Models (VLMs) to assess GUI state consistency with a focus on functionally relevant interaction targets. 
Our framework consists of three stages: interactive region detection, region of interest selection, and attention-driven state comparison.


\subsubsection{Interactive Region Detection}
\label{sec:phase2-1}

Unlike GUI dumped using tools such as UIAutomator~\cite{web:uiautomator}, GUI frames from recordings are plain images and do not contain view hierarchy or accessibility metadata. Therefore, we first identify potential interaction regions from the visual GUI frames.
Note that rather than aiming to identify exact GUI elements, we focus on interactive regions, e.g., coarse-grained areas of the interface that likely contain actionable content. 
This design choice follows a coarse-to-fine paradigm, where coarse localization first narrows the search space, allowing for finer semantic reasoning.

\begin{figure}
	\centering 
	\includegraphics[width=0.9\linewidth]{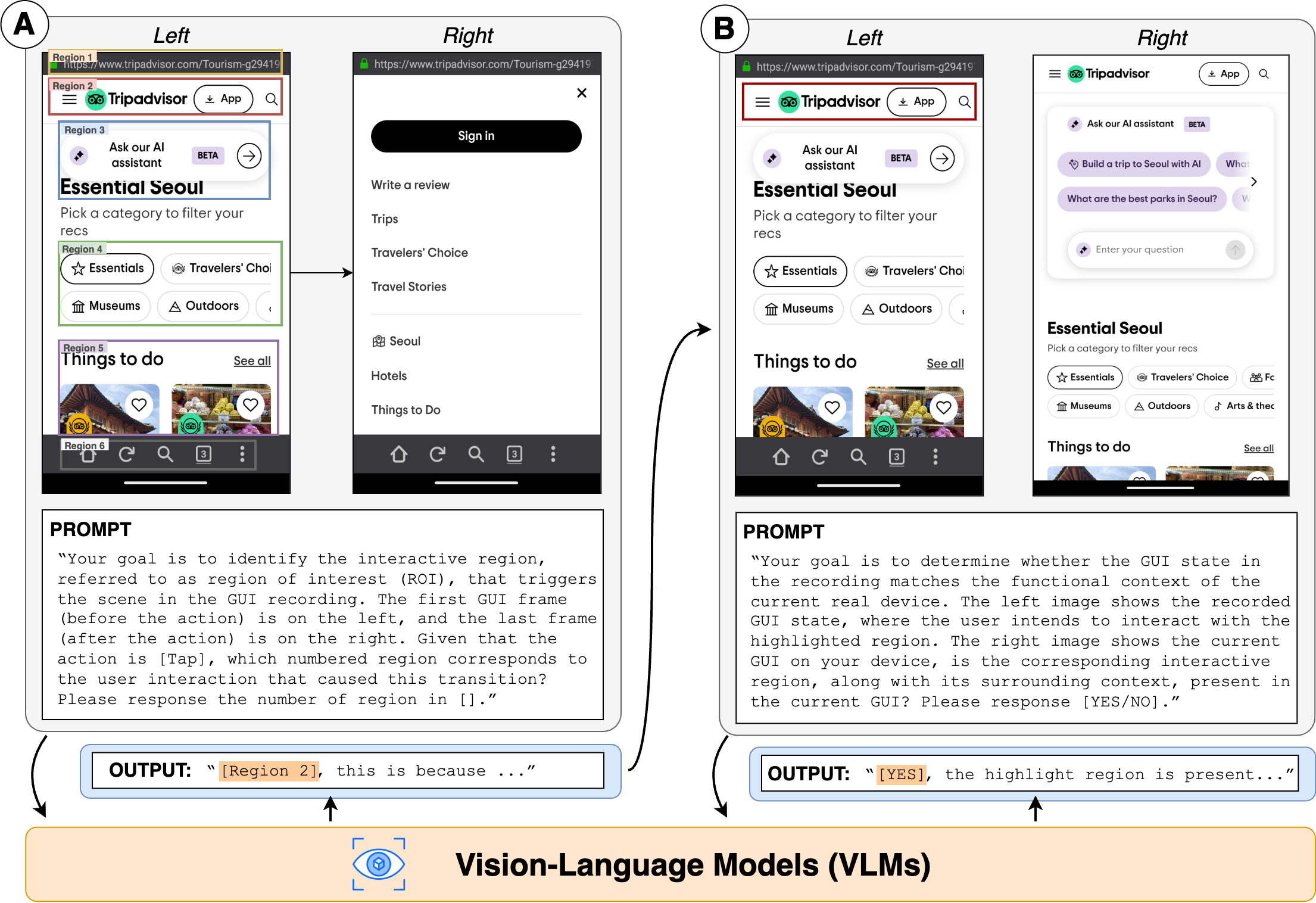} 
	\caption{The example of prompting GUI state comparison.} 
	\label{fig:prompt1}
\end{figure}

To detect these regions, we use GroundingDINO~\cite{liu2024grounding}, a state-of-the-art open-vocabulary object detector, which supports flexible, language-guided detection, making it well-suited for the diverse and dynamic nature of GUI interactive region detection.
In detail, GroundingDINO contains three core components: a vision backbone, a language encoder, and a cross-modal decoder.
The vision backbone is typically a transformer-based architecture pre-trained on large-scale image datasets. It extracts multi-scale visual features from the input GUI screenshot, capturing both global layout and fine-grained GUI element structures.
The language encoder processes textual prompts that describe generic GUI interaction targets, such as ``search bar'', ``layout'', ``button'', or ``text field''. This encoder is based on CLIP, converting the textual queries into high-dimensional embeddings that semantically align with visual regions in the image.
Finally, the cross-modal decoder then integrates visual and textual representations using attention mechanisms. It predicts bounding boxes and associated confidence scores for regions in the GUI that match the semantics of the interactive region. 
This process yields a set of candidate interactive regions from the frame of the action scene.
These candidates serve as anchors for the subsequent reasoning and comparison steps. 
An example of interactive regions localized by GroundingDINO is shown in Fig.~\ref{fig:prompt1}-A.

\subsubsection{Region of Interest Selection}
\label{sec:phase2-2}
From the detected candidate regions, we identify the Region of Interest (ROI), e.g., the specific area corresponding to the user action. 
Accurate ROI selection is critical to focus the comparison on relevant GUI elements while avoiding interference from unrelated GUI dynamics or background elements.

To enhance context, we incorporate temporal information by analyzing both the first and last frames of the action scene, representing the GUI before and after the user action, respectively.
These two frames are merged into a dual-view composite image, preserving the full sequence context. 
Each candidate region from the first frame is labeled (e.g., ``Region 1'', ``Region 2'', etc.) and evaluated within this composite view.
We then prompt the VLMs to identify which candidate region most likely corresponds to the user interaction by formulating a structured query as shown in Fig.~\ref{fig:prompt1}-A.
By grounding the reasoning in both spatial layout and semantic transitions, the VLMs select the true ROI.
The output is a numeric index (e.g., \textit{``Region 2''}), the potential region interacted with during the action scene.





\subsubsection{Attention-Driven State Comparison}
\label{sec:phase2-3}

The final stage is to compare the GUI state captured in the action scene with the current GUI on the device. 
This comparison assesses whether the current interface is functionally consistent with the recorded state, ensuring that the next user action can be meaningfully and reliably replayed.
In detail, we avoid traditional pixel-wise comparison and instead adopt a context-preserving, attention-guided approach using VLMs.
The input to this prompt consists of two GUI screenshots: (1) the first frame of the action scene, in which the selected ROI is visually emphasized (e.g., by drawing a bounding box), and (2) the current GUI screen on the device where replay is intended. 
Both GUI screens are retained to preserve the entire interface context, allowing the model to reason about local alignment (region-level) while maintaining awareness of global consistency (screen-level).
To facilitate the comparison, we design carefully crafted prompts that guide the VLM’s attention toward the intended interaction target and contextual cues from the surrounding interface.
An example prompt is shown in Fig.~\ref{fig:prompt1}-B.
To support interpretability and downstream control, we use binary-response formatting (\textit{``[YES/NO]''}) in prompts, enabling confidence filtering based on the VLM’s response probability.





\subsection{Bug Replay on Device}
\label{sec:phase3}
Once GUI state consistency has been verified, we proceed to replay the recorded user action on the current device. 
If the current GUI state is deemed functionally equivalent to the recorded one, particularly with respect to the identified Region of Interest (ROI), the user action can be directly executed on the current interface.
However, environmental differences (e.g., GUI content, layout shifts, or transient overlays) may cause divergence between the recorded and current states. 
In such cases, direct replay is infeasible, and guided exploration is required to bring the GUI into alignment with the expected precondition.
The overarching goal remains the same: to navigate the current device screen, either with direct guidance from target actions or through exploration without explicit context, in order to ultimately reproduce the bug. 


A key challenge is that the VLMs may lack domain-specific knowledge, that is, they do not possess an inherent understanding of how to execute mobile GUI-level actions. 
To bridge this gap, we design a prompt engineering technique to enable the VLMs to determine the next optimal action based on three core inputs: the executable action space, current GUI perception, and a structured prompt for intended interaction inference.
Over successive reproduction, the VLMs iteratively navigates, adapts, and eventually replays the recorded action sequence to complete the bug replay process.

\renewcommand{\arraystretch}{1}
\begin{table}
    \footnotesize
    \centering
    \caption{Action space.}
    \label{tab:action}
    \begin{tabularx}{\linewidth}{l|l|X} 
        \hline
        \rowcolor{darkgray} \bf{Action}  & \bf{Primitive} & \bf{Description} \\
        \hline
          tap & [tap] [element/back] & click on the element on the GUI screen or system backward to the previous screen. \\
          \hline
          scroll & [scroll] [direction] & navigate the screen vertically, up or down. \\
            \hline
            input & [input] [element] [value] & enter the value into  GUI element field. \\
            \hline
            end & [end] & finish the replay task. \\
            \hline
    \end{tabularx}
\end{table}

\subsubsection{Action Space}
To define the executable capabilities of the VLMs, we formalize a core action space in Table~\ref{tab:action}.
This space includes three commonly used atomic actions: tap, scroll, and input. 
While more complex gestures such as pinch or multi-finger swipes do exist, they are relatively rare in practice. 
For clarity and generality, this work focuses on the most frequently observed actions.
Each action is represented using a structured format tailored to its interaction context, requiring distinct primitives to specify the necessary components. 
For example, the ``tap'' action requires either a target interactive element within the GUI (e.g., a button), represented as [tap] [element], or a system-level backward navigation, represented as [tap] [backward].
Similarly, the scroll action requires a directional specification (e.g., upward or downward), which we represent as [scroll] [direction]. 
When it comes to the ``input'' action, it involves entering a specified value into a particular field, and we structure this as [input] [element] [value]. 
Additionally, we introduce a special action, [end], which signals the completion of the replay sequence and denotes the end of the bug reproduction process.

\subsubsection{Current GUI Perception}

Accurate perception of the current GUI screen is essential for VLMs to understand the environment state.
To this end, we construct a multimodal GUI representation by combining both visual capture and structural metadata.
That is, we first acquire a visual screenshot of the current GUI screen using the Android Debug Bridge (ADB) screencast functionality~\cite{web:adb}, which provides a pixel-level rendering of the interface. 
Simultaneously, we extract the view hierarchy metadata through Android UIAutomator~\cite{web:uiautomator}, which exposes the underlying XML structure of the GUI. 
This metadata contains detailed information about GUI components, including their types, positions, text content, visibility status, and interaction attributes (e.g., clickable, editable).

To extract relevant GUI elements from the XML structure, we perform a depth-first search (DFS) traversal on the view hierarchy tree. 
Starting from the root node, we iteratively explore each child node, resulting in a set of atomic GUI elements suitable for action grounding.
Then, we apply a set-of-mark annotation process~\cite{yang2023set}, in which each detected element is visually marked on the screenshot (e.g., with bounding boxes and labels). 
This augmented image, enriched with both visual and structural cues, serves as the perceptual input.

\begin{figure}
	\centering 
	\includegraphics[width=0.7\linewidth]{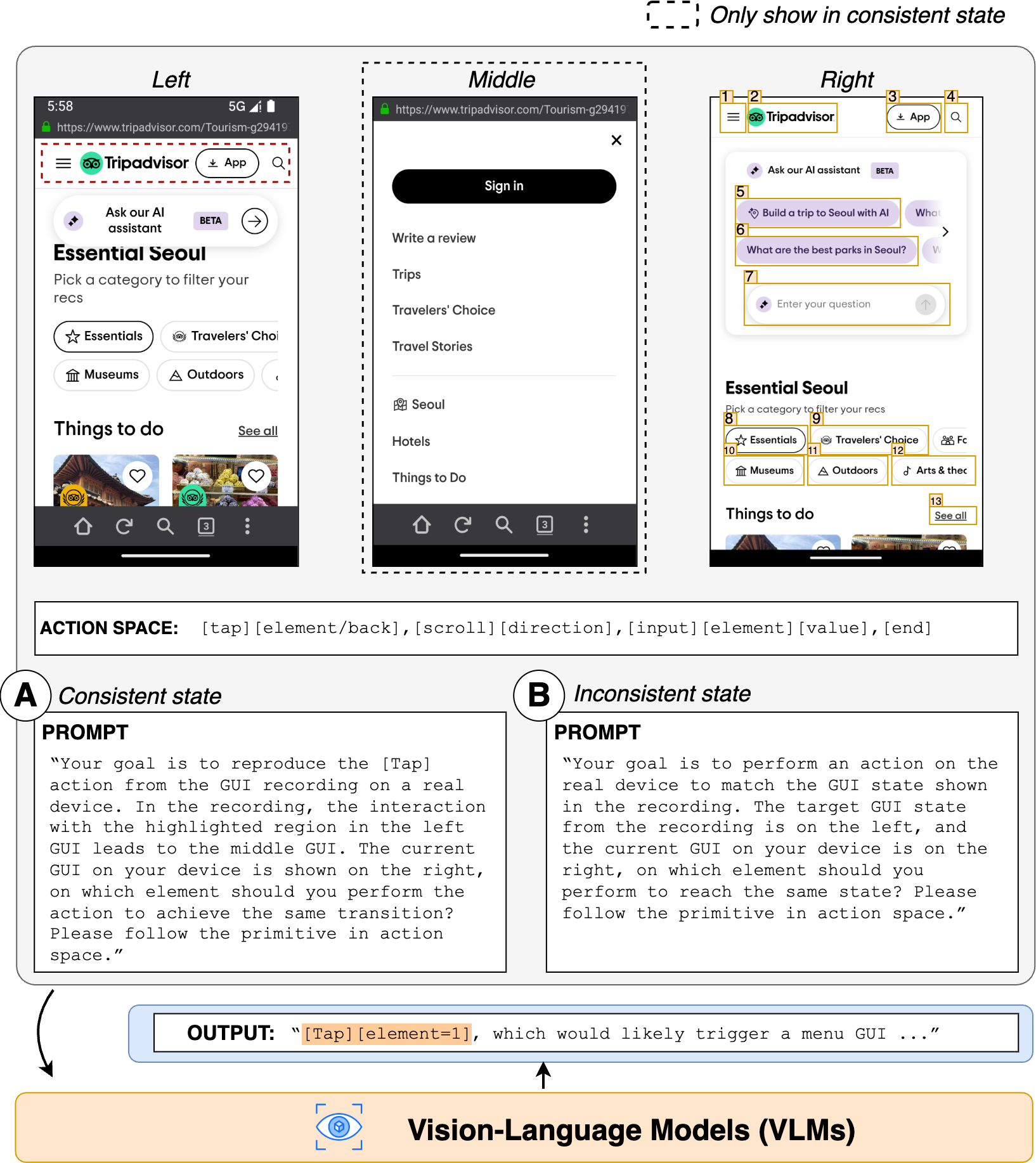} 
	\caption{The example of prompting bug replay on device.} 
	\label{fig:prompt2}
\end{figure}

\subsubsection{Action Inference}
\label{sec:phase3-2}
With the action space defined and the current GUI state perceived, we now leverage these inputs to construct structured prompts that guide the VLMs in inferring the action.
This action inference process is inherently conditioned on the result of the prior GUI state consistency check (Section~\ref{sec:phase2}). 
Accordingly, we adopt a dynamic prompting strategy that tailors the action inference logic based on whether the current GUI state is consistent or inconsistent with the recorded action scene.

\textit{(a) Consistent State}:
When the current GUI matches the first frame of the recorded action scene, the objective is to faithfully replay the user action.
To facilitate this, we construct a context-rich prompt that includes: (i) the first frame with the previously selected ROI highlighting the interaction target (Fig.~\ref{fig:prompt2}-Left), (ii) the last frame of the scene representing the post-action state (Fig.~\ref{fig:prompt2}-Middle), (iii) the current GUI screen (Fig.~\ref{fig:prompt2}-Right), and (iv) the potential action (e.g., tap, scroll, input) obtained from action boundary segmentation in Section~\ref{sec:phase1}.
This composite prompt (as shown in Fig.~\ref{fig:prompt2}-A) provides both temporal and spatial context, enabling the VLMs to infer the most plausible action that progresses the interface toward the intended postcondition.


\textit{(b) Inconsistent State}:
When the current GUI state does not match the first frame of the recorded action scene, we should guide the interface to a compatible state before replaying the original user action. 
In this case, we construct a prompt that includes: (i) the first frame of the action scene, which serves as the target precondition (Fig.~\ref{fig:prompt2}-Left), and (ii) the current GUI screen reflecting the present environment state (Fig.~\ref{fig:prompt2}-Right).
The VLMs is prompted to infer the most effective exploratory action that would transition the interface toward alignment with the recorded state, enabling the continuation of the replay sequence.
An example prompt is shown in Fig.~\ref{fig:prompt2}-B.


This process of inference and execution continues iteratively until the VLMs determine that the bug has been successfully reproduced, with the signal of the [end] action.

\subsection{Implementation}
Our \tool is implemented as a fully automated tool for replaying bug within mobile app based on a GUI recording.
We leverage GroundingDINO~\cite{liu2024grounding} to perform object detection on GUIs throughout the replay process.
We employ GPT-4o, OpenAI’s foundational CUA model, as the core VLM reasoning engine in \tool.
To facilitate automated interpretation of the model’s output, we specify a structured output format and parse the responses using regular expressions (e.g., patterns matching bracketed fields such as [...]).
The GUI structure is extracted using Android UIAutomator~\cite{web:uiautomator}.

\section{Evaluation}
\label{sec:evaluation}
In this section, we describe the procedure we used to evaluate \tool in terms of its performance automatically.
Since our approach consists of three main phases, we evaluate each phase of \tool, including Action Boundary Segmentation (Section~\ref{sec:phase1}), GUI State Comparison (Section~\ref{sec:phase2}), and Bug Replay on Device (Section~\ref{sec:phase3}).

\begin{itemize}[leftmargin=0.3cm] 
    \item \textbf{RQ1:} How accurate is our approach in segmenting the actions from GUI recordings?
    \item \textbf{RQ2:} How accurate is our approach in determining functional consistency in GUI states?
    \item \textbf{RQ3:} How effective is our approach in replaying the bug on device?
    \item \textbf{RQ4:} What is the runtime overhead of our approach for replaying a bug?
\end{itemize}

For \textbf{RQ1}, we evaluate the overall performance of our approach in action boundary segmentation and compare it against state-of-the-art baselines.
For \textbf{RQ2}, we assess the effectiveness of our approach in verifying GUI state consistency, determining whether the current on-device GUI matches the recorded state in the action scene.
For \textbf{RQ3}, we examine the capability of our approach to successfully replay bugs on the device.
Since our approach involves multiple VLM calls (GPT-4o, GroundingDino) during the replay process, it is important to evaluate the runtime overhead introduced by these model inference operations in \textbf{RQ4} to understand the practicality of our approach in real-world settings.

\subsection{RQ1: Performance of Action Boundary Segmentation}
\label{sec:rq1}
\subsubsection{Experimental Setup}
\label{rq1:setup}
To answer RQ1, we first evaluate the effectiveness of our approach (Section~\ref{sec:phase1}) in accurately segmenting user action scenes from GUI recordings.
To ensure a diverse and unbiased dataset, we collect recordings from three existing open-source datasets:
(i) the crash bug reproduction dataset from Themis~\cite{su2021benchmarking};
(ii) the evaluation suite of GIFdroid~\cite{feng2022gifdroid};
and (iii) the study on Android GUI recording V2S~\cite{bernal2020translating}.
Since some GUI recordings overlap across these datasets, we remove duplicates originating from the same issue repository. 
In total, we collect a dataset of 75 unique GUI recordings, yielding 834 action scenes, averaging 11.12 scenes per recording.

\subsubsection{Metrics}
\label{rq1:metric}
We evaluate segmentation performance using three standard scene boundary detection metrics: precision, recall, and F1-score. 
A predicted scene boundary is considered correct if it falls within a tolerance window of ±5 frames~\cite{feng2022gifdroid} from the ground-truth boundary.
Precision is defined as the proportion of predicted action scenes that are correctly matched to ground truth scenes:
\( \textit{precision} = \frac{\# \text{Correctly predicted action scenes}}{\# \text{All predicted action scenes}} \).
Recall is defined as the proportion of all action scenes that are correctly detected by the predictions:
\( \textit{recall} = \frac{\# \text{Correctly predicted action scenes}}{\# \text{All action scenes}} \).
F1-score is the harmonic mean of precision and recall, which combine both of the two metrics above:
\( \textit{F1-score} = \frac{2 \times \text{precision} \times \text{recall}}{\text{precision} + \text{recall}} \).

In addition, we evaluate the correctness of the classified action type during action boundary segmentation.
To achieve this, we adopt the commonly-used metric, accuracy, which measures whether the action scenes have action labels that match the ground-truth labels.
For all metrics, a higher value represents better performance.

\subsubsection{Baselines.}
We compare our method against five state-of-the-art approaches, which are widely used for video segmentation, serving as alternatives to CLIP-based feature similarity and covering both traditional visual similarity methods and learning-based models.
\textit{PySceneDetect}~\cite{web:pyscenedetect} is a practical tool that detects scene boundaries in videos based on threshold-based changes in visual content, such as frame-wise histogram differences.
\textit{Hecate}~\cite{song2016click} is a tool developed by Yahoo that estimates frame quality on the image aesthetics and clusters them into scenes.
\textit{TransNetV2}~\cite{soucek2024transnet} is a deep learning model trained on natural videos, using temporal convolutions to detect scene transitions.
\textit{GPT-4o}~\cite{web:chatgpt} evaluate the capability of a strong, general-purpose VLM to infer semantic scene boundaries and action types from GUI recordings.
\textit{GIFDroid}~\cite{feng2022gifdroid} introduces an image-processing method by using the structural similarity index (SSIM) to segment GUI recordings into user actions.


\subsubsection{Results}
\label{rq1:result}
Table~\ref{tab:performance_rq1} presents the overall performance comparison across all baseline methods.
Our approach significantly outperforms all baselines for scene boundary detection, achieving improvements of 7.4\% in precision, 1.2\% in recall, and 4.9\% in F1-score over the best-performing baseline (\textit{GIFdroid}).
Moreover, our approach achieves higher action-type classification accuracy than all baselines, reaching 0.88, 0.93, and 1.00 accuracy for tap, scroll, and input actions, respectively.

\textit{GPT-4o} exhibits a lower performance, with a precision of 0.25, recall of 0.55, and F1-score of 0.34.
While \textit{GPT-4o} demonstrates strong semantic understanding in many tasks, it struggles with frame-level segmentation, primarily due to its reliance on high-level abstraction that lacks the granularity necessary for accurately identifying scene boundaries—particularly when user actions involve minor but functionally significant interface changes.
This limitation is compounded by ambiguities in prompt interpretation and the definition of user action scenes, leading to underperformance in action boundary segmentation tasks.

The deep learning-based baseline, \textit{TransNetV2}, achieves an average precision of 0.33, recall of 0.56, and F1-score of 0.41.
This relatively low performance is likely due to the domain discrepancy between the natural scenes on which \textit{TransNetV2} was trained and the artificial nature of GUI recordings. 
Unlike natural scenes, which contain real-world elements such as humans, animals, and plants, GUI recordings involve static, highly structured content with specific rendering processes that it fails to generalize to effectively. 

\renewcommand{\arraystretch}{1}
\begin{table}
\footnotesize
\tabcolsep=0.225cm
\centering
\caption{Performance comparison for action boundary segmentation.}
\label{tab:performance_rq1}
\begin{tabular}{l|c|c|c|c|c|c} 
\hline
\multirow{2}{*}{\bf{Method}} & \multicolumn{3}{c|}{\bf{Scene boundary detection}} & \multicolumn{3}{c}{\bf{Action type detection}} \\
\cline{2-7}
& \bf{Precision} & \bf{Recall} & \bf{F1-score} & \bf{Tap Acc.} & \bf{Scroll Acc.} & \bf{Input Acc.} \\ 
\hline
PySceneDetect~\cite{web:pyscenedetect} & 0.42 & 0.55 & 0.48 & - & - & - \\
Hecate~\cite{song2016click} & 0.20 & 0.26 & 0.23 & - & - & - \\
TransNetV2~\cite{soucek2024transnet} & 0.33 & 0.56 & 0.41 & - & - & -\\
GPT-4o~\cite{web:chatgpt} & 0.25 & 0.55 & 0.34 & 0.63 & 0.41 & 0.54 \\
GIFdroid~\cite{feng2022gifdroid} & 0.81 & 0.84 & 0.82 & 0.84 & 0.93 & 1.00 \\
\hline
\bf{\tool} & \bf{0.87} & \bf{0.85} & \bf{0.86} & \bf{0.88} & \bf{0.93} & \bf{1.00} \\
\hline
\end{tabular}
\end{table}

\begin{figure}
	\centering 
	\includegraphics[width=0.85\linewidth]{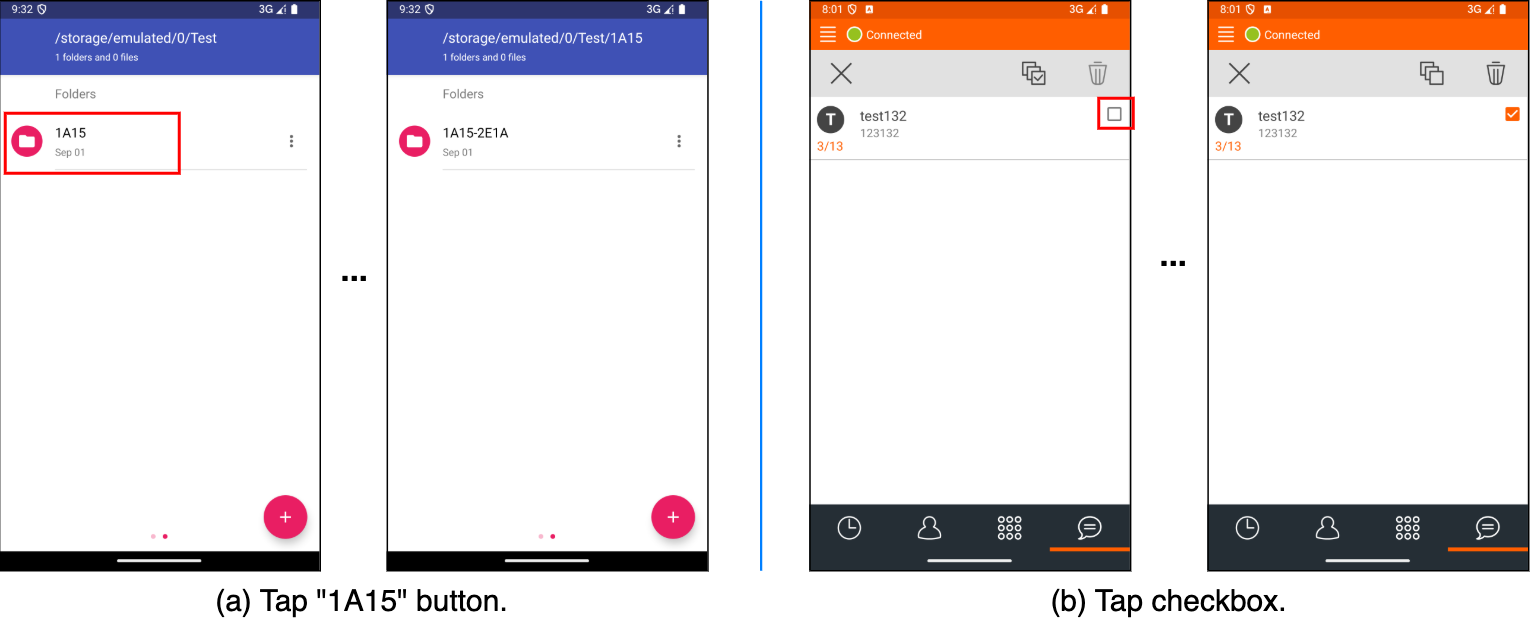} 
	\caption{Examples of failures cases of state-of-the-art baseline GIFdroid in action boundary segmentation.} 
	\label{fig:rq1}
\end{figure}

Traditional image-processing methods, such as \textit{PySceneDetect} and \textit{Hecate}, achieve relatively low F1-scores of 0.48 and 0.23, respectively.
This is because these methods rely on low-level visual heuristics, e.g., histogram differences or aesthetic scoring, that are not well-suited to the subtle and fine-grained transitions present in GUI recordings. 
As a result, they are less robust in handling the visual intricacies of GUIs. 
Among the baselines, \textit{GIFdroid} performs best, achieving 0.81 precision, 0.84 recall, and 0.82 F1-score. 
However, since \textit{GIFdroid} relies on SSIM as a perception metric, it remains limited when segmenting actions that involve large semantic differences but exhibit only minor pixel-level intensity differences.
For example, as illustrated in Fig.~\ref{fig:rq1}, tapping a button triggers a transient overlay transition with low intensity variance: the new screen partially inherits the layout of the previous one, resulting in consecutive frames that appear structurally similar but are semantically different.
In contrast, our approach leverages CLIP-based embeddings, enabling it to capture both semantic and structural changes.

Despite the superior performance, our method still encounters segmentation errors in certain scenarios. 
First, over-segmentation may occur when there are delays in resource loading, e.g., one segment is produced for the GUI transition, and another for the delayed rendering of content triggered by the same user action.
Second, dynamic GUI elements such as advertisements or video playback can produce continuous frame differences that are mistakenly interpreted as separate user actions. 
These challenges highlight opportunities for future improvement, such as integrating motion analysis or adaptive temporal smoothing to reduce false splits.

\subsection{RQ2: Performance of GUI State Comparison}
\label{sec:rq2}
\subsubsection{Experimental Setup}
To address RQ2, we evaluate the effectiveness of our method (as described in Section~\ref{sec:phase2}) in accurately determining functional consistency between the GUI state in each action scene and the current on-device GUI.  Unlike existing datasets that focus on static screen similarity, our evaluation emphasizes functionality-aware comparison, e.g., determining whether the one GUI can support the same user interaction as the other, even if the appearance differs.

To construct the evaluation dataset, we build upon our experimental dataset described in Section~\ref{rq1:setup}, which contains 75 GUI recordings. 
During the reproduction of these recordings on the device, we manually verify whether the action could still be performed and annotate each pair of recorded and current GUI states with a binary label (consistent or inconsistent), which serves as ground truth.
This process yields 463 GUI state comparisons, consisting of 352 consistent and 111 inconsistent cases.


\subsubsection{Metrics}
We adopt the same evaluation metrics as in Section~\ref{rq1:metric}: precision, recall, and F1-score. 
In this context, a prediction is correct if the model's label (consistent or inconsistent) matches the ground-truth annotation for a given GUI state pair. 
These metrics evaluate how accurately the model captures functional alignment between recorded and live GUI states.

\subsubsection{Baselines}
We compare our method against four commonly used image comparison techniques, including one pixel-level (absolute differences \textit{ABS}~\cite{watman2004fast}), one structural-level (\textit{SIFT}~\cite{lowe2004distinctive}), one perceptual-level (\textit{SSIM}~\cite{wang2011ssim}), and one sematic-level (\textit{CLIP}~\cite{radford2021learning}) .
Due to the page limit, we omitted the details of these well-known methods.
Note that these methods only calculate similarity scores without explicit binary classifications of GUI comparison; thus, we adopt a threshold of 0.95 to determine the same GUI state according to the aforementioned experiment and common practice~\cite{feng2022gifdroid,feng2022gifdroid1}.

In addition, we conduct ablation studies to assess the effectiveness of key design components in our approach. Our full approach leverages multiple levels of information, including region-of-interest selection using GroundingDINO on the pre-action frame from the recording, together with the post-action frame and the current GUI captured on the target device, to prompt VLMs for GUI state comparison. We denote this as \textit{Pre-action frame (GroundingDINO) + Post-action frame + Current GUI}.
To isolate the contribution of each component, we evaluate two ablated variants. The first variant, \textit{Pre-action frame + Current GUI}, removes the post-action frame and relies solely on the initial recorded state and the current on-device GUI, allowing us to assess the importance of post-action context. The second variant, \textit{Pre-action frame + Post-action frame + Current GUI}, retains both recorded frames but disables region-level grounding, enabling us to evaluate the impact of incorporating post-action information without explicit region selection.


\renewcommand{\arraystretch}{1}
\begin{table}
\footnotesize
\centering
\tabcolsep=0.3cm
\caption{Performance of GUI state comparison with the baselines.}
\label{tab:performance_rq2}
\begin{tabular}{l|c|c|c} 
\hline
\bf{Method} & \bf{Precision} & \bf{Recall} & \bf{F1-score}  \\ 
\hline
ABS~\cite{watman2004fast} & 0.40 & 0.35 & 0.37 \\
SIFT~\cite{lowe2004distinctive} & 0.52 & 0.50 & 0.51  \\
SSIM~\cite{wang2011ssim} & 0.60 & 0.64 & 0.62 \\
CLIP~\cite{radford2021learning} & 0.61 & 0.53 & 0.57 \\
\hline
\bf{\tool} & \bf{0.86} & \bf{0.88} & \bf{0.87} \\
\hline
\end{tabular}
\end{table}



\subsubsection{Results}
\label{rq2:result}
\begin{figure*}
	\centering 
	\includegraphics[width=0.99\textwidth]{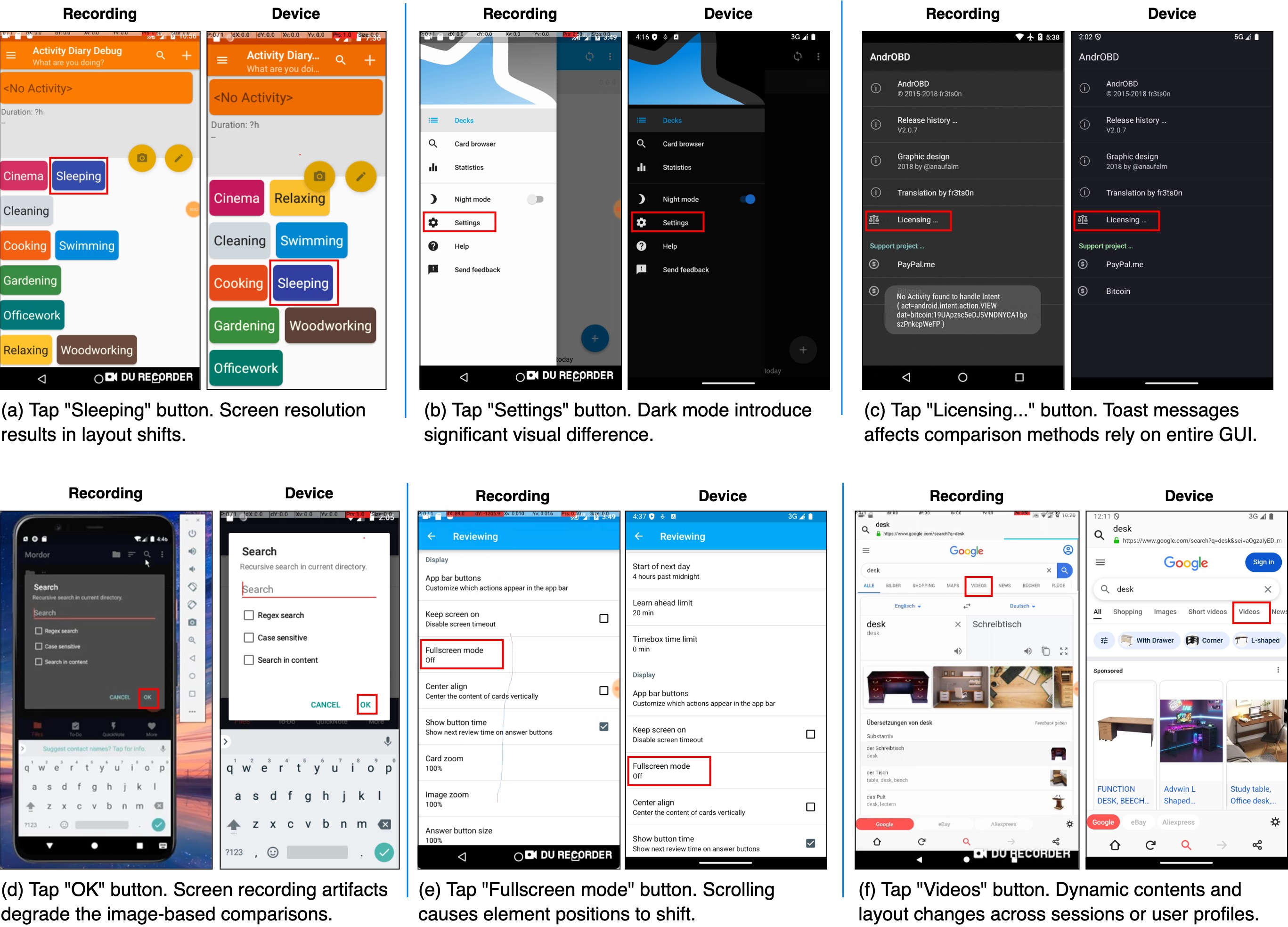} 
	\caption{Examples of failure cases of baselines and ablation studies in GUI comparison.} 
	\label{fig:example_rq2}
\end{figure*}

Table~\ref{tab:performance_rq2} summarizes the performance of the baselines.
Our approach achieves average scores of 0.86 precision, 0.88 recall, and 0.87 F1-score, with a confidence-interval variance below 0.002 across all metrics.
It consistently outperforms all baselines, achieving a 43.3\% improvement in precision, 37.5\% in recall, and 40.3\% in F1-score compared to the best-performing baseline, SSIM. 
Note that we additionally perform a threshold sensitivity analysis for the similarity-based baselines by varying the decision threshold from 0.90 to 0.99. Although the performance varies across this range, the relative ranking of the baselines remains consistent.
These results demonstrate the strength of our functionality-aware, VLM-driven comparison method over traditional visual similarity techniques.
Several factors contribute to this improvement. 
First, GUI environments are dynamic, and differences in screen resolution between the recording and test device can cause layout shifts, which confuse pixel- or structural-based methods such as \textit{ABS} or \textit{SIFT}.
For instance, in Fig.~\ref{fig:example_rq2}(a), small scaling changes result in element misalignment, leading traditional methods to incorrectly classify the states as inconsistent. 
Second, cosmetic settings such as dark mode, font scaling, or language localization (Fig.~\ref{fig:example_rq2}(b)) can introduce significant visual differences without affecting interaction semantics. 
Third, non-functional artifacts such as toast messages or status banners (Fig.~\ref{fig:example_rq2}(c)) may appear during replay, misleading methods that depend on entire GUI matching.
Fourth, screen recording artifacts (e.g., emulator container in Fig.~\ref{fig:example_rq2}(d), watermarks from third-party tools, etc.) can further degrade the performance of image-based comparisons.

\renewcommand{\arraystretch}{1}
\begin{table}
\footnotesize
\centering
\tabcolsep=0.15cm
\caption{Ablation studies of GUI state comparison.}
\label{tab:performance_rq2-2}
\begin{tabular}{l|c|c|c} 
\hline
\bf{Ablation} & \bf{Precision} & \bf{Recall} & \bf{F1-score}  \\ 
\hline
Pre-action frame + Current GUI & 0.65 & 0.60 & 0.62 \\
Pre-action frame + Post-action frame + Current GUI & 0.73 & 0.71 & 0.72 \\
Pre-action frame (GroundingDINO) + Post-action frame + Current GUI & \bf{0.86} & \bf{0.88} & \bf{0.87} \\
\hline
\end{tabular}
\end{table}



Table~\ref{tab:performance_rq2-2} presents the performance of our ablation studies on GUI state comparison. Overall, the full configuration of \tool, \textit{Pre-action frame (GroundingDINO) + Post-action frame + Current GUI}, achieves the best performance, further validating the effectiveness of region-guided prompting.
The \textit{Pre-action frame + Current GUI} variant, which directly compares the pre-action state with the current on-device GUI, performs the worst, achieving an average F1-score of only 0.62.
This poor performance stems from the fact that, without the post-action context, the VLM tends to compare GUIs based on superficial visual attributes (e.g., color schemes, layout density, or background changes) instead of reasoning about whether the depicted state contains an action that can actually be replayed on the current GUI.
Incorporating the post-action frame improves the performance to 0.73 precision, 0.71 recall, and 0.72 F1-score, yielding gains of 8\%, 11\%, and 10\%, respectively. This improvement occurs because the post-action frame provides critical context about the expected outcome of the user interaction, enabling the model to reason beyond static appearance matching. 
However, this variant performs significantly worse in scenarios involving dynamic content, where substantial visual discrepancies appear in the GUI even though the target interactive region remains functionally accessible.
Therefore, incorporating attention-aware guidance over the interactive region of interest in our approach, \textit{Pre-action frame (GroundingDINO) + Post-action frame + Current GUI}, substantially improves performance, e.g., improving 13\%, 17\%, and 15\% for precision, recall, and F1-score, respectively.
For example, in Fig.~\ref{fig:example_rq2}(e), layout shifts caused by scrolling result in position changes of the target ``Fullscreen mode'' button. Similarly, in Fig.~\ref{fig:example_rq2}(f), dynamic GUI changes between sessions alter the surrounding context of the target ``Videos'' tab. 
The ablation baselines fail to recognize these as functional consistent states due to reliance on holistic matching, whereas our region-guided prompt enables robust matching by reasoning over the local interaction semantics.


\subsection{RQ3: Performance of Bug Replay}
\subsubsection{Experimental Setup}
\label{sec:rq3-1}
To answer RQ3, we evaluate the ability of our approach (Section~\ref{sec:phase3}) to effectively replay bugs on real devices using GUI recordings. 
We build upon the dataset described in Section~\ref{rq1:setup}, which originally contains 75 GUI recordings.
However, many recordings are no longer reproducible due to several challenges. 
First, many real-world bugs have already been fixed in newer app versions, and retrieving the exact historical versions required for reproduction is often infeasible. 
Second, some bug scenarios, particularly in financial or social applications, require sensitive credentials, authenticated sessions, or hardware-specific environments beyond the scope of this study. 
Third, a subset of bug recordings lacks visible failure indicators (e.g., crash dialogs or error messages), making it difficult to confirm successful reproduction.
Therefore, we manually replay each GUI recording on a test device to verify that the bug could still be reproduced.
After filtering out unreproducible cases, we obtain a dataset of 38 GUI recordings.
In addition, we manually inspect the GitHub repositories of the apps included in our experimental dataset and collect 6 additional recent GUI recordings.
As a result, our final dataset for RQ3 consists of 44 GUI recordings for experimentation.


\subsubsection{Metrics}
We evaluate performance using two metrics: reproducibility and execution time.
Reproducibility measures whether the approach can successfully execute the user actions in the GUI recording and reach the same buggy state.
A higher score reflects greater robustness in replicating bug-triggering behaviors.
Execution time captures the total runtime of the replay pipeline, including action boundary segmentation, GUI state comparison, and action execution. 
The less time it takes, the more efficiently the method can reproduce the bugs.

\subsubsection{Baselines}
We compare our approach against two state-of-the-art bug replay techniques.
\textit{V2S}~\cite{bernal2020translating} is a visual record-and-replay method that detects user actions by identifying touch indicators in video frames using deep learning and then converting them into replayable scripts.
\textit{GIFdroid}~\cite{feng2022gifdroid} integrates image-processing with static program analysis that maps video keyframes to app states in a UI Transition Graph (UTG), enabling replay by matching visual scenes to graph transitions. 
To support GIFdroid, we first run Droidbot~\cite{li2017droidbot}, an automated UI exploration tool, to collect the required UTG for each target app.
To allow sufficient exploration coverage, we configure multiple exploration time budgets of 0.5, 1, and 2 hours per app.

\subsubsection{Results}
Table~\ref{tab:performance_rq3} summarizes the performance of all methods. 
Our approach achieves an average reproducibility rate of 72.0\% 
within 302.6 seconds for execution time, significantly outperforming both \textit{V2S} and \textit{GIFdroid}.
This improvement is largely due to the robustness of our method against the types of inconsistencies discussed in Section~\ref{rq2:result}, such as resolution mismatches, dynamic UI content, configuration variability, and recording artifacts. 
These factors often cause baseline methods to fail due to their reliance on fragile visual or structural assumptions.

\renewcommand{\arraystretch}{1}
\begin{table}
\footnotesize
\centering
\tabcolsep=0.3cm
\caption{Performance comparison for bug replay.}
\label{tab:performance_rq3}
\begin{tabular}{l|c|c|c} 
\hline
\multicolumn{2}{c|}{\bf{Method}} & \bf{Reproducibility} & \bf{Execution time}  \\ 
\hline
\multicolumn{2}{l|}{V2S~\cite{bernal2020translating}} & 47.7\% & 901.1s \\
\hline
\multirow{3}{*}{GIFdroid~\cite{feng2022gifdroid}} & 0.5h & 45.5\% & 313.5s  \\
\cline{2-4}
 & 1h & 54.5\% & 336.2s \\
\cline{2-4}
 & 2h & 54.5\% & 451.7s \\
\hline
\multicolumn{2}{l|}{\bf{\tool}} & \bf{72.0\%} & \bf{302.6s} \\
\hline
\end{tabular}
\end{table}

In addition, a key advantage of our approach lies in its reduced reliance on auxiliary cues or complete structural knowledge. 
For example, \textit{V2S} requires clearly visible touch indicators to infer user actions, an assumption that rarely holds in user-submitted bug recordings (e.g., from GitHub).
As a result, it fails to handle a substantial portion (21.5\%) of our dataset. 
\textit{GIFdroid}, on the other hand, achieves 47.7\%, 54.5\%, and 54.5\% reproducibility when using 0.5h, 1h, and 2h of automated exploration, respectively. 
The performance gain is largely driven by the increased UTG coverage obtained within the first hour of exploration, after which the improvement quickly plateaus as few to no new states are discovered.
This finding underscores \textit{GIFdroid}’s strong dependence on the completeness and accuracy of the UTG, which in turn introduces several practical limitations observed in our qualitative failure analysis.
First, many GUIs are highly dynamic, and a UTG collected at one point in time can quickly become outdated when the same screen is later rendered with different content (e.g., search results or personalized feeds).
This makes UTG traversal unreliable, for example, the search-result screen in Fig.~\ref{fig:example_rq2}(f) changes substantially across executions, preventing any stable UTG node from consistently representing it. 
Second, the UTG is sensitive to the device configuration used during collection; differences in display resolution, system language, theme, or app-specific settings can alter GUI element positions and the view hierarchy structure. 
As a result, the GUI state observed during replay may not match any UTG node (e.g., Fig.~\ref{fig:difference}).
Third, constructing a UTG for each bug report is time-consuming, limiting the practicality of the approach in real-world settings.
In contrast, our approach is lightweight without requiring instrumentation or app-specific configurations, harnessing the multi-modal reasoning capabilities of vision-language models to achieve functionality-aware GUI matching and robust action inference across diverse device environments and configurations.

We also conduct a text-only bug reproduction baseline, \textit{AdbGPT}, which achieves 41.4\% reproducibility.
While text-based bug reproduction baseline \textit{AdbGPT} is effective for handling structured S2R descriptions, it faces inherent limitations when textual reports omit critical contextual or visual information that is essential for accurate replay. 
For example, a bug report in AndBible\#261~\cite{web:and61} states, ``Trying to do a find in a book, e.g. Josephus, asks you to create an index. When attempting to do so, causes the app to crash''. This description lacks details about the navigation path, intermediate GUI states, or specific interaction steps required to reproduce the bug. 
Similarly, one specific S2R such as ``Click to add an image to a field'' in Anki-Android\#4707~\cite{web:ankidroid4707} is ambiguous when the GUI contains multiple visually identical ``add image'' buttons corresponding to different fields (e.g., Front and Back), whereas a recording clearly disambiguates the intended interaction.
Another example is ``Click the hamburger and select Card Browser'' in Anki-Android\#4977~\cite{web:ankidroid4977}, where interpreting the “hamburger” as the menu icon can introduce semantic uncertainty during automated replay.
These examples illustrate that visual and spatial cues are essential for reliable bug replay but cannot be fully derived from text-only descriptions. GUI recordings, by contrast, retain rich visual context that enables more accurate reproduction.


\begin{figure}
	\centering 
	\includegraphics[width=0.68\linewidth]{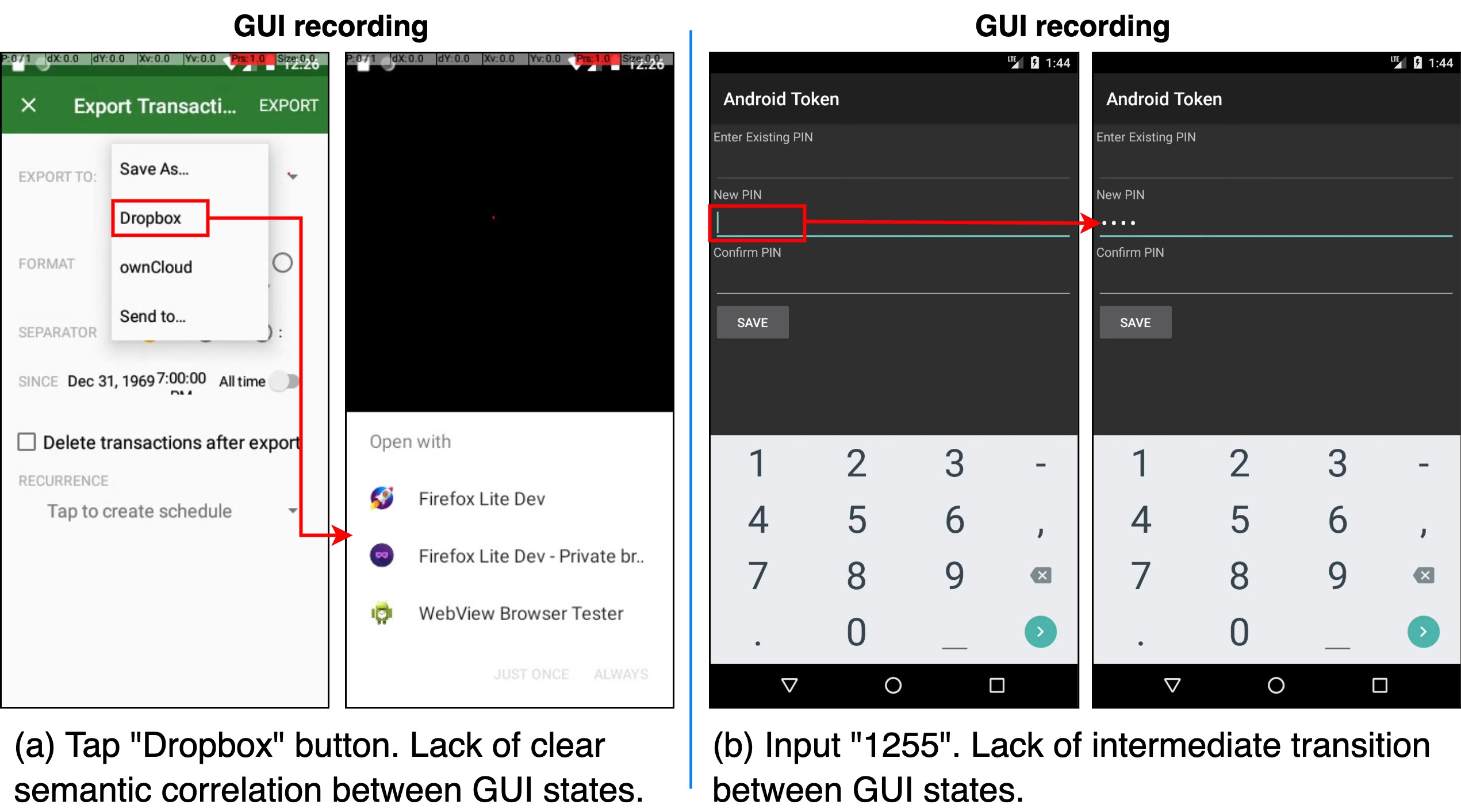} 
	\caption{Examples of failures cases of our approach \tool in bug replaying.} 
	\label{fig:example_rq3}
\end{figure}

Albeit the good performance of our approach, we still fail to replay certain cases. We manually check those failure cases and summarize two common causes.  
First, the lack of clear semantic correlation between GUI states. 
For instance, in Fig.~\ref{fig:example_rq3}(a), tapping the ``Dropbox'' button leads to a next screen with no clear semantic correlation with the original GUI, making it hard for the VLMs to infer the action. 
Second, as shown in Fig.~\ref{fig:example_rq3}(b), inputting a PIN (e.g., ``1255'') results in masked output (``****''), which prevents the VLMs from identifying the precise input without additional context about the intermediate transition.
In the future, we plan to sample intermediate frames within each scene. While this introduces additional computational overhead, it can enrich contextual understanding, particularly in cases involving semantically ambiguous or visually obfuscated transitions. This temporal augmentation may help VLMs more effectively associate cause-effect pairs across challenging UI changes.


\subsection{RQ4: Runtime Overhead}
\subsubsection{Experimental Setup}
To answer RQ4, we measure the runtime overhead associated with VLM calls across different replay phases, including interactive region detection (Section~\ref{sec:phase2-1}), region of interest selection (Section~\ref{sec:phase2-2}), attention-driven state comparison (Section~\ref{sec:phase2-3}), and bug replay action inference (Section~\ref{sec:phase3-2}).
To conduct this evaluation, we use the same dataset from RQ3 in Section~\ref{sec:rq3-1}, which consists of 44 reproducible GUI recordings (38 from the previous datasets and 6 newly collected recordings).

\subsubsection{Metrics}
We evaluate the runtime overhead of our approach using three commonly-used metrics: time latency, token usage, and monetary cost. 
Time latency measures the end-to-end delay of each VLM invocation, including prompt construction, request transmission, and response generation.
Token usage records the number of input and output tokens processed per call, enabling fine-grained analysis of model consumption.
Monetary cost reflects the expense of VLM usage. We compute the total cost based on the official pricing of GPT-4o for input and output tokens at the time of evaluation.

\renewcommand{\arraystretch}{1.1}
\begin{table}
\footnotesize
\centering
\tabcolsep=0.2cm
\caption{Runtime overhead associated with VLM calls.}
\label{tab:performance_rq4}
\begin{tabular}{l|c|c|c|c} 
\hline
\bf{Approach Phase} & \bf{Model} & \bf{Time (s)} & \bf{Tokens (Input / Output)	} & \bf{Cost} \\ 
\hline
Interactive Region Detection & GroundingDINO & 4.17 & – & - \\
Region of Interest Selection & GPT-4o & 4.15 & 2,362 (2,324 / 38) & \$0.005  \\
Attention-Driven State Comparison & GPT-4o & 3.93 & 2,319 (2,318 / 1) & \$0.005  \\
Bug Replay Action Inference & GPT-4o & 5.93 & 3,462 (3,417 / 45) & \$0.008  \\
\hline
\end{tabular}
\end{table}

\subsubsection{Results}
Table~\ref{tab:performance_rq4} presents the runtime overhead associated with each VLM-related component in our approach. For each GUI recording, \tool invokes these components once per action scene.
First, \tool employs Interactive Region Detection using GroundingDINO, which takes an average of 4.17 seconds per scene. Since GroundingDINO is an open-source, locally deployed object detector, it introduces no monetary cost during execution.
Following this step, \tool performs three GPT-4o-based reasoning tasks per action scene: (1) Region of Interest Selection (avg. 4.15s), (2) Attention-Driven State Comparison (avg. 3.93s), and (3) Bug Replay Action Inference (avg. 5.93s). 
We observe moderate input-output token usage across these calls, on average 2,362, 2,319, and 3,462 total tokens (input + output), respectively.
Notably, the Bug Replay Action Inference step consumes more tokens due to the inclusion of multiple GUI frames in the prompt (i.e., the starting frame, post-action frame, and the current on-device GUI), which increases input length.
To mitigate unnecessary token consumption, we incorporate token-saving strategies where applicable. 
For example, in Attention-Driven State Comparison, the model is prompted to respond with a single token (``YES'' or ``NO''), minimizing output cost.
The approximate cost per call is \$0.005 for Region of Interest Selection, \$0.005 for State Comparison, and \$0.008 for Action Inference. 
For example, a typical bug reproduction involving 10 action scenes would incur a total inference cost of around \$0.02.
These results demonstrate that the runtime per VLM call remains within a few seconds, and the total monetary cost per bug replay is low, e.g., typically no more than a few cents. 
This level of efficiency makes \tool practical for integration into developer workflows and continuous integration (CI) pipelines.
Moreover, \tool is model-agnostic, that is, any sufficiently capable open-source or locally deployed VLM can be used in place of GPT-4o.
As open VLMs continue to improve in accuracy and speed, we expect both the cost and latency of \tool to decrease further.

\section{Threats To Validity}

In evaluating our approach, one internal validity arises from the inherent stochasticity of VLMs, which may produce slightly different outputs across runs given the same prompt. 
Consequently, this variability may introduce fluctuations in performance metrics. 
To mitigate this, we executed all VLM-based approaches, including our approach and VLM-related baselines, three times and reported the average performance.
While small variations in outputs were observed, we found that such randomness had limited impact on overall results.
To further understand the effect of VLM variability, we conducted a manual analysis of all failed bug replay attempts during evaluation. 
For each failure, we examined the full replay trajectory, including the sequence of model-generated actions, intermediate screenshots, and the VLM’s textual explanations. We classified a failure as hallucination-related only if the model produced an action clearly inconsistent with the visible GUI, such as referring to a non-existent element or misidentifying an obviously present one.
Based on this analysis, we estimate that hallucination-related errors accounted for $<$2\% of all experiments. 
We therefore believe that such cases do not significantly affect our overall conclusions, and that our approach remains reliable despite the inherent non-determinism of current VLMs.

Another potential source of internal bias stems from the manual labeling of ground-truth in our experimental dataset (Section~\ref{sec:evaluation}).
To ensure annotation reliability, all annotators involved in the labeling process had at least 1.5 years of experience in Android app development and testing.
To further reduce subjectivity, all ground-truth labels were independently annotated and then cross-validated. 
We measured inter-rater reliability using Cohen’s kappa, which yielded a score of 0.87, indicating a high level of agreement. Disagreements were resolved through discussion, and a final random audit was conducted to catch any potential inconsistencies or errors.

As for external validity, the primary threat concerns the size of the dataset for evaluation.
To mitigate this threat, we selected GUI recordings from three prior studies (i.e., V2S~\cite{bernal2020translating}, Themis~\cite{su2021benchmarking}, and GIFdroid~\cite{feng2022gifdroid}) in the domain of GUI recording analysis, which together form a diverse and unbiased dataset. 
In addition, to ensure that our evaluation reflects the current bug-reproduction landscape, we manually collected six additional recent bug recordings by inspecting the GitHub repositories of the apps included in our experimental dataset.
Further empirical studies on larger and more heterogeneous datasets are required to confirm broader generalizability.

\section{Related Work}
A growing body of tools has been dedicated to assisting in recording and replaying bugs.
We review the related work in two main areas: 1) bug record and replay, and 2) vision-language models for software engineering.

\subsection{Bug Record and Replay}
Nurmuradov et al.~\cite{nurmuradov2017caret} introduced a record-and-replay tool for Android applications that captures user interactions by rendering the device screen in a web browser and logging touch events.
It generates heatmaps to visualize user interactions, enabling developers to replay and analyze usage patterns. 
Other tools such as ECHO~\cite{sui2019event}, Reran~\cite{gomez2013reran}, Barista~\cite{ko2006barista}, and WeReplay~\cite{feng2023towards} adopt program analysis techniques to record and replay user actions. 
However, these tools typically require heavy instrumentation or system-level frameworks, such as replaykit~\cite{web:replaykit}, troyd~\cite{jeon2012troyd}, or app-level modifications, which can impose significant overhead and are often too heavy for end users.

Several research have also focused on automating bug replay based on structured bug reports. 
For example, Fazzini et al.~\cite{fazzini2018automatically} proposed Yakusu, which combines program analysis with natural language processing (NLP) to synthesize executable test cases from bug descriptions. 
ReCDroid~\cite{zhao2019recdroid}, introduced by Zhao et al., enhanced this idea by incorporating lexical knowledge to improve crash reproduction accuracy. 
However, these methods often overlook the temporal relationships among reproduction steps (e.g., ``after tapping X, open Y''), which are crucial for accurately modeling interaction flows. 
Liu et al.~\cite{liu2020automated} addressed this limitation by proposing MaCa, an NLP and machine learning-based approach that normalizes S2R (steps-to-reproduce) instructions and extracts structured entities. 
Recent advances in large language models (LLMs) have inspired prompt-based methods for textual bug replay, including AdbGPT~\cite{feng2024prompting} and ReBL~\cite{wang2024feedback}, which use few-shot learning and feedback prompting that significantly improve the performance. 
Despite their advancements, these techniques still heavily depend on the quality and consistency of natural language in bug reports~\cite{erfani2014works,bettenburg2008makes}, including formatting, vocabulary, and granularity.

A few studies~\cite{xie2020uied,xie2022psychologically,feng2024mud,chen2019gallery,feng2022gallery,feng2023video2action, feng2023efficiency, feng2025towards} have leveraged visual information from bug recordings to assist replay. 
For instance, Feng et al.~\cite{feng2023read} proposed CAPdroid, which utilizes computer vision techniques to automatically generate subtitles for bug recordings to support bug reproduction.
Wang et al.~\cite{wang2025empirical} conducted an empirical study on visual information in bug reports to understand its effectiveness in supporting bug comprehension and reproduction.
Bernal et al.~\cite{bernal2020translating} proposed V2S, a video-to-script system that uses deep learning to detect user actions based on touch indicators visible in the recording, translating them into executable test scripts. 
Similarly, Feng et al.~\cite{feng2022gifdroid} introduced GIFdroid, which applies image processing techniques to extract keyframes from screen recordings and map them to states in a GUI transition graph (UTG) for replay. 
In contrast, our approach offers a lightweight solution that does not require touch indicators or UTG collection.
In detail, we leverage vision-language models to reason over both GUI appearance and interaction context, enabling functionality-aware action inference and robust replay across different device environments, making it more broadly applicable and easier to deploy in real-world scenarios.


\subsection{Vision-Language Models for Software Engineering}
Following the success of Large Language Models (LLMs) across a wide range of natural language processing tasks, researchers have increasingly explored their application to software engineering. 
Early efforts have primarily focused on code-related tasks~\cite{deng2022fuzzing,huang2023api}, such as automated code generation, completion, and translation.
A notable example is Codex~\cite{web:codex}, the model behind GitHub Copilot, which demonstrated the ability to translate natural language comments into executable code, effectively reducing developer workload and narrowing the gap between user intent and implementation.

More recently, research has begun shifting toward multi-modal models, particularly Vision-Language Models (VLMs), which integrate visual and textual modalities to support richer forms of reasoning across tasks~\cite{liu2024right,dai2024gpt4ego}, such as visual question answering, human action recognition, and embodied robotics.
Despite their success in these areas, applications of VLMs to software engineering remain relatively nascent.
One emerging direction lies in GUI-based software tasks, such as generating executable GUI code from screenshots.
For example, Wan et al.~\cite{wan2024mrweb} introduced MRWeb, which systematically explored prompting strategies, including zero-shot and chain-of-thought prompting, to improve contextual understanding and generation accuracy.
Building on this idea, Wan et al.~\cite{wan2025divide} proposed DCGen, a pipeline that segments GUI screenshots into components and then prompts VLMs to generate corresponding code snippets, which are reassembled into full-page implementations.
Recently, self-refinement prompting has been investigated to further improve model outputs: models are instructed to generate a candidate solution, critique it, and revise iteratively. 
For instance, Zhou et al.~\cite{zhou2025declarui} proposed DeclarUI, which enhances GUI code generation quality through this refinement loop.

Beyond GUI code generation, VLMs have also unlocked new opportunities in GUI automation and software testing, where they are increasingly explored for their ability to perceive, reason about, and interact with user interfaces.
Claude~\cite{web:computeruse}, for instance, introduced the concept of a ``computer use agent'', a class of agents that operate directly over GUIs using vision-language reasoning rather than API hooks.
Extending this concept, frameworks such as CogAgent~\cite{hong2024cogagent}, WebLINX~\cite{lu2024weblinx}, and UX-LLM~\cite{pourasad2024does} apply VLMs to automate GUI navigation, execute predefined tasks, and conduct usability testing in web or desktop environments.
These systems represent some of the earliest attempts to exploit the perceptual and semantic capabilities of VLMs for GUI interaction tasks that were previously dominated by heuristic or API-level methods.
Additionally, Feng et al.~\cite{feng2026smart} recently proposed a taxonomy of Autonomous GUI Automation Levels, offering a conceptual framework to categorize GUI agents based on their degree of autonomy and intelligence.

Despite these promising developments, most existing approaches are limited to reasoning over single static GUI screenshots, which restricts their ability to model temporal dynamics and user workflows that unfold over time.
In contrast, our work is the first to leverage VLMs in the context of GUI recordings, treating them as sequences of visual frames to reason about user intent and interaction across time. 
By enabling both bug interpretation and replay from screen recordings, our approach extends the scope of VLMs beyond static perception to temporal, functionality-aware GUI understanding, bridging a critical gap in automated debugging and offering a new paradigm for VLM-based software testing.

\section{Conclusion and Future Work}
GUI recordings in video-based bug reports are becoming increasingly popular due to their ease of creation and rich contextual information. 
This paper introduces \tool, a lightweight and fully automated approach for reproducing bugs on devices using GUI recordings.
Unlike prior methods that depend on pixel-level similarity, explicit touch indicators, or pre-collected UI transition graphs, \tool leverages CLIP-based embeddings together with the semantic reasoning capabilities of Vision-Language Models (VLMs) like GPT-4o to understand both visual context and user intent over time. 
In detail, we first segment GUI recordings into discrete scenes representing individual user interactions. We then propose a novel attention-guided prompting strategy combined with region-level reasoning to compare the GUI state between each recorded scene and the current device screen. Based on the comparison, our approach infers and executes the appropriate sequence of actions to robustly guide the bug replay process.
Through extensive experiments, \tool significantly outperforms state-of-the-art baselines and ablation variants, demonstrating its effectiveness in accurately reproducing bugs.

While \tool demonstrates the feasibility of leveraging VLMs for interpreting and replaying GUI recordings, several avenues remain open for future exploration. 
First, we plan to enhance action inference by incorporating intermediate animation frames between UI transitions, rather than relying solely on pre- and post-action frames. This would provide richer temporal context and improve inference accuracy. 
Second, we aim to expand the action space supported by our approach to include more complex high-level interactions, such as long-press, swipes, pinch-in gestures (pinch/zoom), and multi-touch inputs, thereby broadening the applicability of \tool to a wider range of real-world mobile apps and interaction scenarios.
Finally, our results suggest that combining mature off-the-shelf tools (e.g., OCR, CLIP, GroundingDINO, and GPT-4o) can yield an effective solution for visual bug reproduction. As these underlying tools continue to advance, we expect \tool to benefit accordingly and plan to investigate such improvements in future work.


\section{Data Availability}
The source code is available at \url{https://github.com/sidongfeng/ViBR} for future research.

\begin{acks}
We sincerely appreciate the support of Jannis Arnold in conducting experiments.
This work was partially supported by OpenAI Researcher Access Program and Amazon Research Award.
Tingting Yu was supported by the U.S. National Science Foundation (NSF) under grant CCF-2403747 and Aldeida Aleti was supported by the Australian Research Council under grant DP210100041.
\end{acks}

\bibliographystyle{ACM-Reference-Format}
\bibliography{main}

\end{document}